\documentclass[10pt]{article}
\usepackage{latexsym}
\usepackage{amssymb}
\usepackage{amsmath}
\usepackage{amscd}
\usepackage{amsthm}
\usepackage[left=2cm,top=2.5cm,right=2.5cm,bottom=1.5cm]{geometry}

\usepackage[dvips]{graphicx}
\usepackage{hyperref}

\begin{document}
\begin{center}
\Large{\bf{LRS Bianchi-I Anisotropic Cosmological Model with
Dominance of Dark Energy}} \vspace{10mm}

\normalsize{Anil Kumar Yadav$^\dag$ and Bijan Saha$^\ddag$}\\ \vspace{4mm}
\normalsize{$^\dag$Department of Physics, Anand Engineering
College, Keetham, Agra-282 007, India} \\
\vspace{2mm}
\normalsize{$^\dag$E-mail: abanilyadav@yahoo.co.in}\\
\vspace{4mm}
\normalsize{$^\ddag$Laboratory of Information Technologies, Joint Institute for Nuclear Research \\
Dubna - 141980, Russia} \\
\vspace{2mm} \normalsize{$^\ddag$E-mail : bijan@jinr.ru}
\normalsize{$^\ddag$URL : http://bijansaha.narod.ru}
\end{center}
%%\date{}
%%\makefile
\begin{abstract}
The present study deals with spatially homogeneous and anisotropic
locally rotationally symmetric (LRS) Bianchi type I cosmological
model with dominance of dark energy. To get the deterministic model
of Universe, we assume that the shear scalar $(\sigma)$ in the model
is proportional to expansion scalar $(\theta)$. This condition leads
to $A=B^{n}$, where $A$,\;$B$ are metric potential and $n$ is
positive constant. It has been found that the anisotropic
distribution of dark energy leads to the present accelerated
expansion of Universe. The physical behavior of the
Universe has been discussed in detail.\\
\end{abstract}
%%%%%%%%%%%%%%%%%%%%%%%%%%%%%%%%%%%%%%%%%%%%%%%%%%%%%%%%%%%%%%%%%%%%%%%%%%%%%%%%%%%

%%%%%%%%%%%%%%%%%%%%%%%%%%%%%%%%%%%%%%%%%%%%%%%%%%%%%%%%%%%%%%%%%%%%%%%%%%%%%%%%%%%
%%%%%%%%%%%%%%%%%%%%%%%%%%%%%%%   SECTION 1  %%%%%%%%%%%%%%%%%%%%%%%%%%%%%%%%%%%%%%
\section{Introduction}
The discovery of the accelerated mode of expansion of the Universe
stands as a major breakthrough of the observational cosmology.
Survey of cosmological distant type Ia supernovae (SNe Ia; Riess et
al 1998; Perlmutter et al 1999) indicated the presence of a new
unaccounted-for Dark energy (DE) that opposes the self-attractions
of matter and causes the expansion of Universe to accelerate. This
acceleration is realized with negative pressure and positive energy
density that violate the strong energy condition. This violation
gives a reverse gravitational effect. Due to this effect, the
Universe gets a jerk and the transition from the earlier
deceleration phase to the recent acceleration phase  takes place
(Caldwell et al. 2006). The cause of this sudden transition and the
source of accelerated expansion are still unknown. The state of the
art in cosmology has led to the following present distribution of
the energy densities of the Universe: $4\%$ for baryonic matter,
$23\%$ for
non baryonic dark matter and $73\%$ so-called DE (Spergel et al. 2007).\\

The isotropy of the cosmic microwave background (CMB) radiation,
first seen by the cosmic background explorer (COBE) satellite (Smoot
et al. 1992) and then reinforced by the Wilkinson Microwave
Anisotropy Probe (WMAP) data (Hinshaw et al. 2003), together with
the assumption that we are not in spacial position in the Universe,
underlines the cosmological principles, according to which we live
in a homogeneous and isotropic Universe described by a FRW
line-element. Tiny deviation from perfect isotropy at the level of
$10^{-5}$, have also been reported by Bennett et al (1996) and
thereafter confirmed by high resolution WMAP data. The observed CMB
anisotropy spectrum is in impressive agreement with the predictions
of $\Lambda$CDM model. Koivisto and Mota (2008a, 2008b) proposed the
mechanism of DE with anisotropic equation of state (EoS) parameter
which is very attractive because cosmic anisotropy originates from
the actual dominant component of the Universe and then could be
directly tested, for example, by either observations of the
magnitude and redshift of type Ia supernovae or cosmic parallax
effects of the distance source. DE has been conventionally
characterized by the equation of state (EoS) parameter
$\omega^{(de)}=p^{(de)}/\rho^{(de)}$ which is not necessarily
constant. The simplest DE candidate is the vacuum energy $(\omega =
-1)$, which is argued to be equivalent to the cosmological constant
$(\Lambda)$ (Martins, 2002). However, it is well known, there are
two difficulties arising from the cosmological constant scenario,
namely the two famous cosmological constant problems - the $fine\;
tuning$ and the $cosmic\; coincidence$ one. An alternative proposal
is the concept of dynamical DE. Such a scenario is often realized by
some scalar field mechanism and suggests that the energy form with
negative pressure is provided by a scalar field evolving under a
properly constructed potential. So far, a large class of
scalar-field DE models have been studied, including quintessence
$viz \;\omega^{(de)}>-1$ (Steinhardt et al 1999), phantom $viz\;
\omega^{(de)}<-1$ (Caldwell 2002) and quintom (that can across from
phantom region to quintessence region). The quintom scenario of DE
is designed to understand the nature of DE with $\omega^{(de)}$
across $-1$ (Setare 2006). Recently, Cai et al (2010), Setare and
Saridakis (2007, 2009b) have studied the DE models with EoS
parameter across -1 which give a concrete theoretical justification
for quintom paradigm. In addition, the other proposals on DE include
interacting DE model (Setare 2007) and braneworld model (Setare and
Saridakis 2009a) etc. By combining data from seven CMB experiments
with large scale structure data, the Hubble parameter measurement
from the Hubble space-telescope and luminosity measurements of SN
Ia, Melchiorri et al (2003) demonstrated
the bound on $\omega^{(de)}$ to be $-1.38 < \omega^{(de)} < -0.82$ at $95 \%$ confidence level.\\

The simplest of anisotropic models are Bianchi type-I homogeneous
models whose spatial sections are flat but the expansion or
contraction rate are direction dependent. For studying the possible
effects of anisotropy in the early Universe on present day
observations many researchers (Huang 1990; Chimento et al. 1997;
Lima and Troden 1996; Lima and Maia 1994; Pradhan and Singh 2004;
Pradhan and Pandey 2006; Saha 2006a, 2006b) have investigated
Bianchi type-I models from different point of view. The binary
mixture of perfect fluid and DE has been studied for
Bianchi type I (Saha 2005). Some Authors (Akarsu and Kilinc 2010;
Yadav and Yadav 2011; Yadav et al 2011b; Kumar and Yadav 2011; Amirhashchi et al
2011 and recently Yadav 2011) have studied anisotropic DE models
with constant
deceleration parameter (DP).\\

In this paper, we considered minimally interacting perfect fluid and
DE components with proportionality relation between shear scalar and
expansion within the framework of LRS Bianchi-I space-time in
general relativity. The paper is organized as follows: In section 2,
the models and field equations have been presented. Section 3 deals
with the exact solutions of field equations and physical behavior of
the model. The statefinder and distance modulus curves are described
in section 4. Finally the
results are discussed in section 5.\\
%%%%%%%%%%%%%%%%%%%%%%%%%%%%%%%%%%%%%%%%%%%%%%%%%%%%%%%%%%%%%%%%%%%%%%%%%%%%%%%%%%%
%%%%%%%%%%%%%%%%%%%%%%%%%%%%%%%  SECTION 2  %%%%%%%%%%%%%%%%%%%%%%%%%%%%%%%%%%%%%%%%
\section{The Metric and Field  Equations}
We consider the LRS Bianchi type I metric of the form
\begin{equation}
\label{eq1}
ds^2 = -dt^2 + A^2dx^2 + B^2 \left(dy^2 + dz^2\right) \;,
\end{equation}
where, A and B are functions of $t$ only. This ensures that the model is
spatially homogeneous.\\

The Einstein's field equations in case of a mixture of perfect fluid
and DE components, in the units $8\pi G=c=1$,  read as
\begin{equation}\label{eq2}
R^{i}_{\;j}-\frac{1}{2} g^{i}_{\;j}R =- T^{i}_{\;j},
\end{equation}
where $T^{i}_{\;j}=T^{(m)\;i}_{\;\;j}+T^{(de)\;i}_{\;\;j}$ is the overall energy
momentum tensor with $T^{(m)}_{ij}$ and $T^{(de)}_{ij}$ as the
energy momentum tensors of ordinary matter and DE, respectively.
These are given by
\begin{equation}\label{eq3}
T^{(m)\;i}_{\;\;j}=diag[-\rho^{(m)},p^{(m)}\;,p^{(m)},p^{(m)}] \;,\\
                   =diag[-1,\omega^{(m)},\omega^{(m)},\omega^{(m)}]\rho^{(m)}\;
\end{equation}
and
\begin{equation}\label{eq4}
T^{(de)\;i}_{\;\;j}=diag[-\rho^{(de)},p^{(de)},p^{(de)},p^{(de)}]\\
                   =diag[-1,\omega^{(de)},\omega^{(de)},\omega^{(de)}]\rho^{(de)}\;
\end{equation}
where $\rho^{(m)}$ and $p^{(m)}$ are, respectively the energy
density and pressure of the perfect fluid component or ordinary
baryonic matter while $\omega^{(m)}=p^{(m)}/\rho^{(m)}$ is its EoS
parameter. Similarly,  $\rho^{(de)}$ and $p^{(de)}$ are,
respectively the energy density and pressure of the DE component
while $\omega^{(de)}=p^{(de)}/\rho^{(de)}$ is the corresponding EoS
parameter.

The Einstein's field equations (\ref{eq2}) for the line-element (\ref{eq1})
lead to the following system of equations
\begin{equation}
\label{eq5}
2\frac{\ddot{B}}{B} + \frac{\dot{B}^2}{B^2}  = -\omega^{(m)}\rho^{(m)}-\omega^{(de)}\rho^{(de)}\;,
\end{equation}
\begin{equation}
\label{eq6}
\frac{\ddot{A}}{A} + \frac{\ddot{B}}{B} + \frac{\dot{A}\dot{B}}{AB} = -\omega^{(m)}\rho^{(m)}-\omega^{(de)}\rho^{(de)} \;,
\end{equation}
\begin{equation}
\label{eq7}
\frac{\dot{B}^2}{B^2} + 2\frac{\dot{A}\dot{B}}{AB}
 = \rho^{(m)}+\rho^{(de)}\;.
\end{equation}
The Bianchi identity $G^{\; ij}_{\; ;j} =0$ yields
\begin{equation}\label{eq8}
\dot{\rho}^{(m)}+3(1+\omega^{(m)})\rho^{(m)}H+\dot{\rho}^{(de)}+3(1+\omega^{(de)})\rho^{(de)}H=0,
\end{equation}
with $H$ being the mean Hubble parameter, which for LRS Bianchi I
space-time can be defined as
\begin{equation}
\label{eq11}
H=\frac{\dot{a}}{a}=\frac{1}{3}\left(\frac{\dot{A}}{A}+2\frac{\dot{B}}{B}\right)\;
\end{equation}
Here, and in what follows, over-dots indicates differentiation with
respect to $t$ and $a$ is the average scale factor of LRS Bianchi
type I model:
\begin{equation}
 \label{eq9}
a=(AB^{2})^{\frac{1}{3}}.
\end{equation}
The spatial volume (V) is given by
\begin{equation}
\label{eq10}
V = a^{3} = AB^{2}.
\end{equation}

The expansion scalar ($\theta$), shear scalar ($\sigma$) and mean anisotropy parameter ($A_{m}$) are defined as
\begin{equation}
\label{eq12}
\theta =3H = \frac{\dot{A}}{A}+2\frac{\dot{B}}{B}\;,
\end{equation}
\begin{equation}
\label{eq13}
 \sigma^{2}=\frac{1}{2}\left(\sum_{i=1}^{3} H_{i}^{2}-\frac{1}{3}\theta^{2}\right)\;,
\end{equation}
\begin{equation}
\label{eq14} A_{m} =
\frac{1}{3}\sum_{i=1}^{3}\left(\frac{H_{i}-H}{H}\right)^{2}.
\end{equation}

%%%%%%%%%%%%%%%%%%%%%%%%%%%%%%%%%%%%%%%%%%%%%%%%%%%%%%%%%%%%%%%%%%%%%%%%%%%%%%%%%%%%%%%%%%%%%%%%%%
%%%%%%%%%%%%%%%%%%%%%%%%%%%%%%%  SECTION 3  %%%%%%%%%%%%%%%%%%%%%%%%%%%%%%%%%%%%%%%%%%%%%%%%%%%%%
\section{Solutions of the Field Equations}
In order to solve the field equations completely, firstly we assume
that the perfect fluid and DE components interact minimally.
Therefore, the energy momentum tensors of the two sources may be
conserved separately.

The energy conservation equation $T^{(m)~ij}_{~;j} =0$, of the
perfect fluid leads to
\begin{equation}\label{eq15}
\dot{\rho}^{(m)}+3(1+\omega^{(m)})\rho^{(m)}H=0\;,
\end{equation}
whereas the energy conservation equation
$T^{(de)~ij}_{~;j} =0$, of the DE component
yields
\begin{equation}\label{eq16}
\dot{\rho}^{(de)}+3(1+\omega^{(de)})\rho^{(de)}H=0\;.
\end{equation}

Following Akarsu and Kilinc (2010), we assume that the EoS
parameter of the perfect fluid to be a constant, that is,
\begin{equation}\label{eq17}
\omega^{(m)}=\frac{p^{(m)}}{\rho^{(m)}}=const.,
\end{equation}
while $\omega^{(de)}$ has been allowed to be a function of time
since the current cosmological data from SNIa, CMB and large scale
structures mildly favor dynamically evolving DE crossing the phantom divide line (PDL)
as discussed in Section 1.

Finally, we constrain, the system of equation with proportionality
relation between shear $(\sigma)$ and expansion $(\theta)$. This
condition leads to the following relation between the metric
potentials
\begin{equation}
\label{eq18}
A =  B^{n}
\end{equation}
where $n$ is positive constant. For anisotropic model $n \ne 1$.\\
Equations (\ref{eq6}), (\ref{eq7}) and (\ref{eq18}) lead to
\begin{equation}
\label{eq19}
\frac{\ddot{B}}{B}+(n+1)\frac{\dot{B}^{2}}{B^{2}} = 0\;.
\end{equation}
The solution of equation (\ref{eq20}) is given by
\begin{equation}
\label{eq20}
B = (k_{1}t+k_{0})^{\frac{1}{n+2}}\;,
\end{equation}
Where $k_{0}$ and $k_{1}$ are the constants of integration.\\
From equations (\ref{eq18}) and (\ref{eq20}), we obtain
\begin{equation}
\label{eq21} A = (k_{1}t+k_{0})^{\frac{n}{n+2}}\;.
\end{equation}
The rate of expansion in the direction of $x$, $y$ and $z$ are given by
\begin{equation}
\label{eq22}
H_{x} = \frac{\dot{A}}{A} = \frac{nk_{1}}{(n+2)}\frac{1}{(k_{1}t+k_{0})}\;,
\end{equation}
\begin{equation}
\label{eq23} H_{y} = H_{z} = \frac{\dot{B}}{B} =
\frac{k_{1}}{(n+2)}\frac{1}{(k_{1}t+k_{0})}\;.
\end{equation}
The mean Hubble's parameter $(H)$, expansion scalar $(\theta)$ and shear scalar $(\sigma)$ are given by
\begin{equation}
\label{eq24}
H=\frac{k_{1}}{3(k_{1}t+k_{0})}\;,
\end{equation}
\begin{equation}
\label{eq25}
\theta=\frac{k_{1}}{(k_{1}t+k_{0})}\;,
\end{equation}
\begin{equation}
\label{eq26} \sigma^{2} =
\frac{(n-1)^{2}k_{1}^{2}}{3(n+2)^{2}}\frac{1}{(k_{1}t+k_{0})^{2}}\;.
\end{equation}
The spatial volume (V), mean anisotropy parameter $(A_{m})$ and DP $(q)$ are found to be
\begin{equation}
\label{eq27}
V = (k_{1}t+k_{0})\;
\end{equation}
\begin{equation}
\label{eq28}
A_{m} = \frac{2(n-1)^{2}}{(n+2)^{2}}\;
\end{equation}
\begin{equation}
\label{eq30} q=\frac{d}{dt}\left(\frac{1}{H}\right)-1=2\,.
\end{equation}
From equations (\ref{eq25}) and (\ref{eq26}), we obtain
\begin{equation}
\label{eq29} \frac{\sigma}{\theta}=\frac{(n-1)}{\sqrt{3}(n+2)}\,.
\end{equation}

It is important to note here that the proportionality relation
between shear and expansion leads to the positive deceleration
parameter $(q)$ with isotropic distribution of DE in LRS Bianchi -I
space-time. Since we are looking for a model explaining an expanding
Universe with acceleration, so, we assume the anisotropic
distribution of DE to ensure the present acceleration of Universe.
Thus equations (\ref{eq5}), (\ref{eq6}) and (\ref{eq16}) may be
re-written as
\begin{equation}
\label{eq31}
2\frac{\ddot{B}}{B} + \frac{\dot{B}^2}{B^2}  = -\omega^{(m)}\rho^{(m)}-
(\omega^{(de)}+\delta)\rho^{(de)}\;,
\end{equation}
\begin{equation}
\label{eq32}
\frac{\ddot{A}}{A} + \frac{\ddot{B}}{B} + \frac{\dot{A}\dot{B}}{AB} =
-\omega^{(m)}\rho^{(m)}-(\omega^{(de)}+\gamma)\rho^{(de)}\;,
\end{equation}
\begin{equation}
\label{eq33}
\dot{\rho}^{(de)}+3\rho^{(de)}(1+\omega^{(de)})H + \rho^{(de)
}(\delta H_{x}+2\gamma H_{y}) = 0\;.
\end{equation}
The third term of equation (\ref{eq33}) arises due to the deviation
from $\omega^{(de)}$ while the first and second terms of equation
(\ref{eq33}) are deviation free part of $T^{(de)\;i}_{\;\;j}$.
According to equation (\ref{eq33}), the behavior of $\rho^{(de)}$ is
controlled by the deviation free part of of EoS parameter of DE but
deviation will affect $\rho^{(de)}$ indirectly, since as can be seen
later, they affect the value of EoS parameter. But we are looking
for physically viable models of Universe consistent with
observations. Hence we constrained $\delta(t)$ and $\gamma(t)$ by
assuming the special dynamics which is
consistent with (\ref{eq33}). The dynamics of skewness parameter on
x-axis $(\delta)$ and y-axis or z-axis $(\gamma)$ are given by
\begin{equation}
\label{eq34}
\delta=-\frac{2mHH_{y}}{\rho^{(de)}}\;,
\end{equation}
\begin{equation}
\label{eq35} \gamma=\frac{mHH_{x}}{\rho^{(de)}}\;,
\end{equation}
where $m$ is the dimensionless constant that parameterizes the amplitude of the
deviation from $\omega^{(de)}$ and can be given real values.\\
Now, subtracting equation (\ref{eq32}) from (\ref{eq33}), we get
\begin{equation}
\label{eq36}
\frac{\ddot{B}}{B}-\frac{\ddot{A}}{A}+\frac{\dot{B}^{2}}{B^{2}}
-\frac{\dot{A}\dot{B}}{AB}=(\gamma-\delta)\rho^{(de)}\;,
\end{equation}
Using equations (\ref{eq11}), (\ref{eq18}), (\ref{eq34}) and
(\ref{eq35}), from (\ref{eq36}) we obtain
\begin{equation}
\label{eq37}
\frac{\ddot{B}}{B}+\left[\frac{3(n^{2}-1)+m(n+2)^{2}}{3(n-1)}\right]\frac{\dot{B}^{2}}{B^{2}}=0\;,
\end{equation}
The general solution of equation (\ref{eq37}) has the form
\begin{equation}
\label{eq38}
B = (k_{1}t+k_{0})^{\frac{3(n-1)}{n_{1}}}\;,
\end{equation}
where $n_{1}= 3 N_1 + m(n+2)^{2}$ with $N_1 = (n-1)(n+2).$\\
For $A$ in this case we find
\begin{equation}
\label{eq39}
A = (k_{1}t+k_{0})^{\frac{3n(n-1)}{n_{1}}}\;.
\end{equation}
It is important to note here that we obtain power law solution by assuming
proportionality relation between shear scalar$(\sigma)$ and expansion
$(\theta)$ which seems to describe the dynamics of Universe from big
bang to present epoch while a series of works: Yadav and Yadav (2011); Yadav et al (2011);
Amirhashchi et al (2011); Kumar and Yadav (2011); Akarsu and Kilinc (2010) and
recently Yadav (2011) have obtained the power law solution by assuming special
law of variation of  Hubble's parameter. So, we represent the new features of
power law expansion. In this paper, we show how $\sigma\propto\theta$ model
with metric (\ref{eq1}) behaves in presence of perfect fluid and anisotropic DE components.\\
Now, the metric (\ref{eq1}) reduces to
\begin{equation}
\label{eq40}
ds^{2}=-dt^{2}+(k_{1}t+k_{0})^{\frac{6n(n-1)}{n_{1}}}dx^{2}+(k_{1}t+k_{0})^{\frac{6(n-1)}{n_{1}}}(dy^{2}+dz^{2})\;.
\end{equation}
In view of the assumption $\omega^{(m)} = const.$, equation (\ref{eq15}) can be integrated to obtain
\begin{equation}
\label{eq41}
\rho^{(m)}=\rho_{0}a^{-3(\omega^{(m)}+1)}\;,
\end{equation}
where $\rho_{0}$ is the positive constant of integration.\\
The physical parameter such as directional Hubble parameter $(H_{x},
H_{y} \;or\; H_{z})$, average Hubble parameter $(H)$, anisotropy
parameter $(A_{m})$, shear scalar $(\sigma)$, expansion scalar
$(\theta)$ and spatial volume $(V)$ of model (\ref{eq40}) are
respectively given by
\begin{equation}
\label{eq42}
H_{x} = \frac{3n(n-1)}{n_{1}}\frac{k_{1}}{(k_{1}t+k_{0})}\;,
\end{equation}
\begin{equation}
\label{eq43}
H_{y}=H_{z}=\frac{3(n-1)}{n_{1}}\frac{k_{1}}{(k_{1}t+k_{0})}\;,
\end{equation}
\begin{equation}
\label{eq44} H=\frac{N_1}{n_{1}}\frac{k_{1}}{(k_{1}t+k_{0})} \;,
\end{equation}
\begin{equation}
\label{eq45} A_{m}=\frac{2(n-1)^{2}}{(n+2)^{2}}\;,
\end{equation}
\begin{equation}
\label{eq46}
\sigma^{2}=\frac{3(n-1)^{4}k_{1}^{2}}{n_1^2(k_{1}t+k_{0})^{2}}\;,
\end{equation}
\begin{equation}
\label{eq47}
\theta=\frac{3N_1}{n_{1}}\frac{k_{1}}{(k_{1}t+k_{0})}\;,
\end{equation}
\begin{equation}
\label{eq48} V = (k_{1}t+k_{0})^{(3N_1/n_{1})}\;.
\end{equation}
The average scale factor $(a)$ and DP $(q)$ are found to be
\begin{equation}
\label{eq49} a=(k_{1}t+k_{0})^{(N_1/n_{1})}\;,
\end{equation}
\begin{equation}
\label{eq50} q=\frac{n_1}{N_1}-1 = 2 + \frac{m}{N_1}(n+2)^2\;.
\end{equation}
As one sees, the DP $q$ is a constant. The sign of $q$ indicates
whether the model inflates or not. A positive sign of $q$, i. e.
$n_1/N_1 > 1$ corresponds to standard decelerating model whereas
negative sign of $q$, i. e. $0 < n_1/N_1< 1$ indicates acceleration.
The recent observations SN Ia, reveal that the present Universe is
accelerating and the value of DP lies somewhere in the range $-1 < q
< 0$. It follows that in the derived model, one can choose the value
of
DP consistent with observations.\\

From (\ref{eq41}) we than find the energy density of perfect fluid
\begin{equation}
\label{eq53} \rho^{(m)}= \rho_0(k_1 t + k_0)^{-3(\omega^{(m)} +
1)N_1/n_1}\;,
\end{equation}
on account of (\ref{eq53}) from (\ref{eq7}) we obtain the dark
energy density as
\begin{equation}
\label{eq54}
\rho^{(de)}=\frac{9(2n+1)(n-1)^{2}k_{1}^{2}}{n_{1}^{2}(k_{1}t+k_{0})^{2}}-
\frac{\rho_{0}}{(k_{1}t+k_{0})^{3(\omega^{m}+1)N_1/n_1}}\;,
\end{equation}

Now from (\ref{eq34}) and (\ref{eq35}) skewness parameter are
obtained as
\begin{equation}
\label{eq51} \delta(t) = -\frac{6m(n-1)N_1
k_{1}^{2}}{9(2n+1)(n-1)^{2}k_{1}^{2}-\rho_{0}n_{1}^{2}(k_{1}t+k_{0})
^{2-3(\omega^{(m)}+1)N_1/n_1}}\;,
\end{equation}

\begin{equation} \label{eq52} \gamma(t)=\frac{3mn(n-1) N_1
k_{1}^{2}}{9(2n+1)(n-1)^{2}k_{1}^{2}-\rho_{0}n_{1}^{2}(k_{1}t+k_{0})
^{2 - 3(\omega^{(m)}+1)N_1/n_1}}\;,
\end{equation}

The EoS parameter of DE is given by
\begin{equation}
\label{eq55} \omega^{(de)}=
-\frac{\frac{\omega^{(m)}\rho_{0}}{(k_{1}t+k_{0})^{3(\omega^{(m)}+1)N_1/
n_1}}+\frac{3(n-1)^{2}k_{1}^{2}(6-2m(n+2)-n_{1}/(n-1))}{n_{1}^{2}(k_{1}t+k_{0})^{2}}}{\frac{9(2n+1)
(n-1)^{2}k_{1}^{2}}{n_{1}^{2}(k_{1}t+k_{0})^{2}}-\frac{\rho_{0}}{(k_{1}t+k_{0})^{3(\omega^{m}+1)N_1/n_1}}}\;,
\end{equation}
%%%%%%%%%%%%%%%%%%%%%%%%%%%%%%%%%%

\begin{figure}
\begin{center}
\includegraphics[width=4.0in]{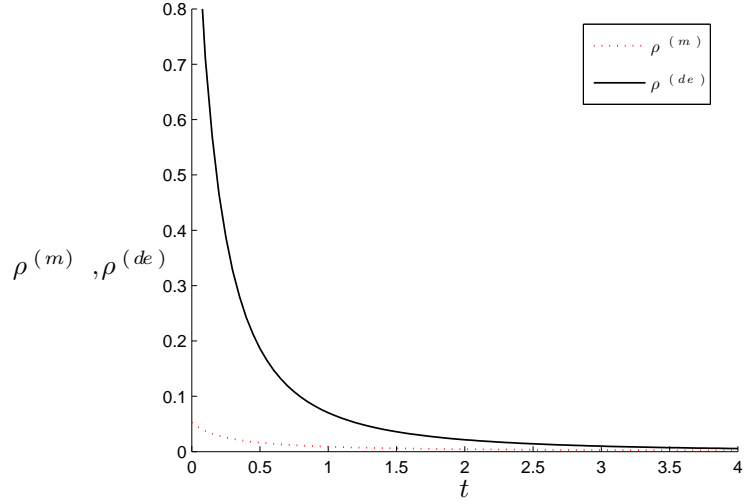}
\caption{Plot of matter density $(\rho^{(m)})$ and DE density
$(\rho^{(de)})$ versus time.} \label{fg:saha2.eps}
\end{center}
\end{figure}

\begin{figure}
\begin{center}
\includegraphics[width=4.0in]{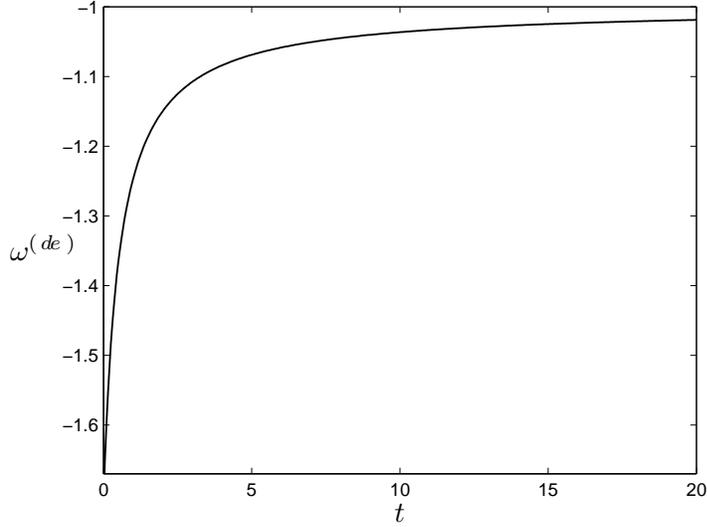}
\caption{Plot of EoS parameter of DE components $(\omega^{(de)})$
versus time.} \label{fg:saha3.eps}
\end{center}
\end{figure}
It is observed that at $t = -\frac{k_{0}}{k_{1}}$, the spatial
volume vanishes while all other parameters diverge. Thus the derived
model starts expanding with big bang singularity at $t =
-\frac{k_{0}}{k_{1}}$ which can be shifted to $t=0$ by choosing
$k_{0}=0$. This singularity is point type because the directional
scale factors $A(t)$ and $(B(t))$ vanish at initial moment. From
$\textbf{Fig. 1}$, we observe that $\rho^{(m)}$ as well as
$\rho^{(de)}$ remains positive during the cosmic evolution.
Therefore the weak energy condition (WEC) as well as null energy
condition (NEC) are obeyed in the derived model. Further
$\rho^{(m)}$ and $\rho^{(de)}$ decrease with time, and approach to a
small positive values at the present epoch. The parameter $H$,
$\sigma$ and $\theta$ start off with extremely large values and
continue to decrease with expansion of universe. $\textbf{Fig. 2}$
clearly shows that $\omega^{(de)}$ evolves with negative and it's
range is in nice agreement with large scale structure data (Komatsu
et al 2009).\\
\begin{figure}
\begin{center}
\includegraphics[width=4.0in]{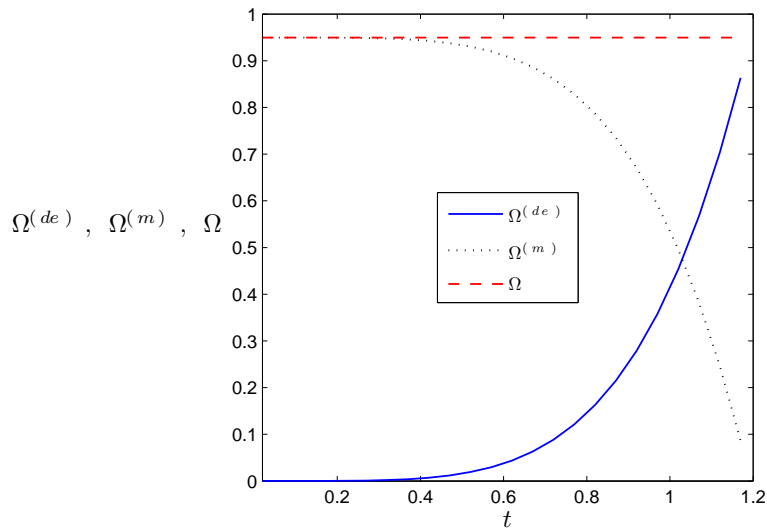}
\caption{Plot of density parameters versus time.}
\label{fg:saha4.eps}
\end{center}
\end{figure}
The density parameters of perfect fluid and DE are as follows:
\begin{equation}
\label{eq56} \Omega^{(m)}=\frac{\rho_{0}n_{1}^{2}}{3N_1^2k_{1}^{2}}
(k_{1}t+k_{0})^{2 - 3(\omega^{(m)}+1)N_1/{n_{1}}}\;
\end{equation}
\begin{equation}
\label{eq57}
\Omega^{(de)}=\frac{3(2n+1)}{(n+2)^{2}}-\frac{\rho_{0}n_{1}^{2}}{3
N_1^2 k_{1}^{2}}(k_{1}t+k_{0})^{2 -3(\omega^{(m)}+1) N_1/{n_{1}}}\;
\end{equation}
Adding equation (\ref{eq56}) and (\ref{eq57}), the overall density
parameter $(\Omega)$ is obtained as
\begin{equation}
\label{eq58}
\Omega=\Omega^{(m)}+\Omega^{(de)}=\frac{3(2n+1)}{2(n-1)^{2}}A_{m}
\end{equation}
This shows that the overall density parameter $(\Omega)$ depends on
the anisotropy parameter $(A_{m})$. \textbf{Fig. 3} demonstrates the
behavior of density parameters in the evolution of Universe with
appropriate choice of constants of integration and other physical
parameters using reasonably well known situations. We observe that
initially the ordinary matter density dominates the Universe. But
later on, the DE density dominates the evolution which is probably
responsible for the accelerated expansion of present-day Universe.

%%%%%%%%%%%%%%%%%%%%%%%%%%%%%%%%%%%%%%%%%%%%%%%%%%%%%%%%%%%%%%%%%%%%%%%%%%%%%%%%%%%%%%%%%%%%%%%%%%
%%%%%%%%%%%%%%%%%%%%%%%%%%%%%%%  SECTION 4  %%%%%%%%%%%%%%%%%%%%%%%%%%%%%%%%%%%%%%%%%%%%%%%%%%%%%
\section{The statefinder \& Distance Modulus Curves}

Sahni et al (2003) proposed a cosmological diagnostic pair $\{r,
s\}$ called state finder, which is defined as
\begin{equation}
\label{eq59}
r=\frac{\dddot{a}}{aH^{3}}=\left(1-\frac{n_{1}}{N_1}\right)\left(1-\frac{2n_{1}}{N_1}\right)\;,
\end{equation}
\begin{equation}
\label{eq60}
s=\frac{r-1}{3(q-\frac{1}{2})}=\frac{\left(1-\frac{n_{1}}{N_1}\right)\left(1-\frac{2n_{1}}{N_1}
\right)-1}{3\left(\frac{n_{1}}{N_1}-\frac{3}{2}\right)}\;.
\end{equation}
The dynamics of statefinder $\{r,s\}$ depends on constant $n_{1}$ and
$N_{1}$. It follows that in derived model, one can choose the pair
of statefinder which can successfully differentiate between a wide
variety of DE models including cosmological constant, quintessence,
phantom, quintom, the chaplygin gas, braneworld models and
interacting DE models. For example if we put $n_{1}=0$, the
statefinder pair will be $\{1, 0\}$  which yields the $\Lambda CDM $
(cosmological constant cold dark matter) model. The statefinder
diagnosis for holographic DE model in non flat
Universe has been analyzed by Setare et al (2007).\\

The distance modulus is given by
\begin{equation}
\label{eq61} \mu = 5\;log\;d_{L} +25\;,
\end{equation}
where the luminosity distance $d_{L}$ is defined as
\begin{equation}
\label{eq62} d_{L} = r_{1}(1+z)a_{0}\;,
\end{equation}
where $z$ and $a_{0}$ represent red shift parameter and
present scale factor respectively.\\
Let us now assume that $T = k_{1}t+k_{0}$. Thus equation
(\ref{eq49}) may be rewritten as
\begin{equation}
\label{eq63} a=T^{n_{2}}\;,
\end{equation}
where $n_{2} =N_1/n_{1}$\\
For determination of $r_{1}$, we assume that a photon emitted by a
source with co-ordinate $r=r_{1}$ and $T=T_{1}$ and received at a
time $T_{0}$ by an observer located at $r=0$. Then we determine
$r_{1}$ from
\begin{equation}
\label{eq64} r_{1}=\int_{T_{1}}^{T_{0}}\frac{dT}{a}\;.
\end{equation}
Solving equations (\ref{eq61})$-$(\ref{eq64}), one can easily obtain
the expression for distance modulus $(\mu)$ in term of red shift
parameter $(z)$ as
\begin{equation}
\label{eq65} \mu =
5\;log\;\left[\frac{n_{2}k_{1}}{H_{0}(1-n_2)(1+z)^{\frac{1-2n_{2}}{n_{2}}}}\left((1+z)^{\frac{1-n_{2}}{n_{2}}}-1\right)\right]+25\;.
\end{equation}
\begin{figure}
\begin{center}
\includegraphics[width=4.0in]{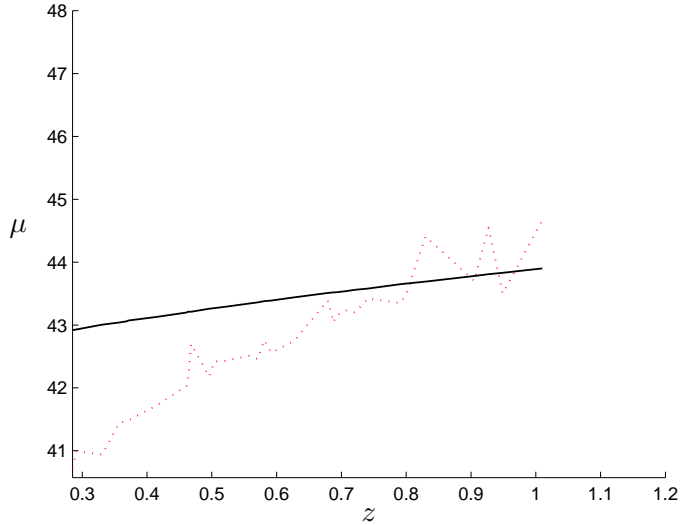}
\caption{Distance modulus as a function of the redshift to the
derived model compared with SNLS type Ia supernovae data from Astier
et al. (2006)} \label{fg:saha1.eps}
\end{center}
\end{figure}
The comparison between the derived model and SNLS type Ia supernovae
data can be seen in $\bf{Fig. 4}$. The dotted line represents the
observed distance modulus by SNLS type Ia supernovae data where as
solid line represents the analyzed distance modulus $\mu$
of the derived model. It is observed that the derived model is best fit with high redshift values.\\
%%%%%%%%%%%%%%%%%%%%%%%%%%%%%%%%%%%%%%%%%%%%%%%%%%%%%%%%%%%%%%%%%%%%%%%%%%%%%%%%%%%%%%%%%%%%%%%%%%
%%%%%%%%%%%%%%%%%%%%%%%%%%%%%%%  SECTION 5  %%%%%%%%%%%%%%%%%%%%%%%%%%%%%%%%%%%%%%%%%%%%%%%%%%%%%
\section{Conclusion}
In this paper, we have studied a spatially homogeneous and anisotropic LRS Bianchi-I space time 
filled with perfect fluid and anisotropic DE possessing dynamical energy density. Studying the 
interaction between the ordinary matter and DE will open up the possibility of detecting DE. 
It should be pointed out that evidence was recently provided by Abell-Cluster A586 in support 
of interaction between DE and dark matter (Bertolami et al 2007; Le Delliou et al 2007). Let us now concentrate on some other works on minimally 
interacting perfect fluid and DE models of Universe, especially the works by Akarsu and Kilinc (2010) 
and Yadav (2011). In both the works, the field equations have been solved by using special law of 
variation of Hubble's parameter which yields the constant value of DP where as the present investigation 
is one with power law solution by taking into account the proportionality relation between shear scalar $(\sigma)$ 
and expansion scalar $(\theta)$. It is to be noted that our procedure of solving the field equations is 
altogether different from what Akarsu and Kilinc (2010) have adapted in LRS Bianchi-I space-time. However, 
a common feature of all these power law solution models is to describe the dynamics of Universe from 
big bang to present epoch.\\       
 
In the derived model, the EoS parameter of DE $(\omega^{(de)})$ is obtained as time varying and 
it is evolving with negative sign which may be attributed
to the current accelerated expansion of Universe. Also note that 
the isotropic distribution of DE is not possible in LRS Bianchi Type I
space-time because the isotropic distribution of DE leads to the
positive value of DP which can not explain the current accelerated
expansion of Universe while for anisotropic distribution of DE, DP
evolves with negative sign. The distance modulus curve of derived model is in good agreement with
SNLS type Ia supernovae for high redshift value which in turn imply 
that the derived model is physically realistic.\\

 The age of Universe is given by
$$ T_{0}=\frac{N_1}{n_{1}}H_{0}^{-1}-\frac{k_{0}}{k_{1}}$$
which is different from the present estimate i.e.
$T_{0}=H_{0}^{-1}=14Gyr$. But if we take $k_{0}=0$ and $N_1/n_{1}=1$, i.e., $m = 2
(1-n)/(2+n)$, where $m$ is the constant describing the anisotropy of DE 
and $n$ is the constant giving the proportionality condition between shear and expansion 
scalar, then the derived model is in good
agreement with the present age of Universe.

%%%%%%%%%%%%%%%%%%%%%%%%%%%%%%%%%%%%%%%%%%%%%
\section*{Acknowledgements}
The authors would like to thank the anonymous referee for his/her useful comments to improve this work 
and drawing our attention to a couple of references relevant to our studies. Author (AKY) is thankful to
 The Institute of Mathematical Science (IMSc), Chennai,
India for providing facility and support where part of this work was
carried out.

%\newline
%\nonumsection{References}

\end{document}